%% file: main.tex
\documentclass[aps,prb,reprint,groupedaddress]{revtex4-2}
\usepackage{amssymb}
\usepackage{amsfonts}
\usepackage{xcolor}

\usepackage{amsmath}
\usepackage{comment}
\usepackage{braket}
\usepackage{booktabs}
\usepackage{graphicx}
\input{macros} 

\begin{document}
\title{Symmetry-enforced agreement of Kohn--Sham and many-body Berry phases in the SSH--Hubbard chain}
\author{Kai Watanabe}
\email{kaiw.sshhrm@gmail.com}
\affiliation{Independent Researcher, Akita, Japan}
\date{\today}

\begin{abstract}
\input{sections/00_abstract}
\end{abstract}
\maketitle
\input{sections/01_intro}
\input{sections/02_model}
\input{sections/03_method}
\input{sections/04_results}

\input{sections/05_discussion}

\newpage
\appendix
\input{sections/A_appendix}

\bibliography{refs}

\end{document}

%% file: macros.tex
\providecommand{\ket}[1]{\lvert #1 \rangle}
\providecommand{\bra}[1]{\langle #1 \rvert}
\providecommand{\braket}[2]{\langle #1 \vert #2 \rangle}



%% file: sections/00_abstract.tex
We study when a density-matching Kohn--Sham (KS) description can reproduce a many-body Berry phase in a correlated insulator, despite the fact that geometric phases are functionals of the wave function.
Focusing on the one-dimensional SSH--Hubbard chain on a ring as a controlled interacting topological model, we introduce a $U(1)$ twist $\theta$ (flux insertion).
The many-body ground state along the full twist cycle is computed by the density-matrix renormalization group (DMRG), while the onsite interaction $U$ is tuned from the noninteracting to the strong-coupling regime.
At half filling in the inversion-symmetric gapped regime, our DMRG calculations show that the density remains constant within numerical accuracy over the entire $(\theta,U)$ range studied.
Thus, the density has no dependence on either the flux $\theta$ or the interaction strength $U$.
Accordingly, the symmetry-preserving density constraint collapses the KS reference to an SSH-type quadratic representative with $U$-independent geometric diagnostics.
Nevertheless, the many-body wave function exhibits a nontrivial geometric response: the quantum metric associated with the $\theta$-parametrized ground-state manifold depends on $\theta$ at intermediate $U$ and is strongly suppressed at large $U$, consistent with the charge fluctuation freezing.
Intriguingly, the KS and many-body Berry phases coincide throughout the gapped regime as $U$ is tuned from weak to strong coupling.
We show that this agreement is best understood as symmetry-enforced $\mathbb{Z}_2$ sector matching, rather than as evidence that the density encodes the many-body Berry connection.

%% file: sections/01_intro.tex
\section{Introduction}
Density functional theory (DFT) in the Kohn--Sham (KS) formulation provides a practical
mapping from an interacting many-electron problem to a noninteracting KS effective system
that reproduces the ground-state density \cite{Hohenberg1964,Kohn1965}.
A basic question, central to correlated topology, is how strongly a \emph{density-matching}
KS description constrains geometric and topological information encoded in the many-body
wave function.

Topological phases are characterized by geometric invariants, most notably Berry phases
and the associated polarization response \cite{Berry1984,KingSmithVanderbilt1993,Zak1989}.
For noninteracting electrons these quantities are determined by the occupied Bloch bands
and can be evaluated directly from a band structure.
In practical materials computations, such band structures are often obtained from
DFT-based electronic-structure calculations, where the Kohn--Sham construction provides
a convenient noninteracting representation.
This connection underlies the widespread use of DFT workflows in band-topological
characterization \cite{Hasan2010,Qi2011}.

However, topological quantities in strongly correlated electron systems are not constrained
in a straightforward manner by the ground-state density alone.
As interactions increase, the ground state becomes genuinely many-body, and geometric
quantities such as polarization and the associated Berry phase are functionals of the
many-body wave function.
Indeed, it has been emphasized that these phases are not uniquely determined, in general,
by reduced information such as single-particle expectation values or the density
\cite{Resta1998,Resta2006,Vanderbilt2009,Vanderbilt2018}.
This raises the possibility that a density-matching KS description fails to reconstruct the
interaction-dependent holonomy.
Nevertheless, there exist model settings in which the KS description appears to reproduce
the same phase in the correlated regime \cite{Resta2018}.

Under what circumstances can a density-matching KS description reproduce the many-body Berry phase in a correlated regime?
To address this question, we study a minimal interacting lattice model, the one-dimensional SSH--Hubbard chain~\cite{Mondal2021HalfSSHHubbard,Le2020SSHH}.
This minimal strongly correlated topological setting allows us to make the relevant conditions explicit in a controlled manner and to identify the underlying mechanisms without additional model-dependent complications.
In this framework, we compare the many-body and density-matching KS Berry phases defined along the same $U(1)$ twist cycle, focusing on how the agreement (or lack thereof) evolves with $U$.

The paper is organized as follows.
In Sec.~\ref{sec:model}, we define the one-dimensional SSH--Hubbard model and introduce the
unit-cell (sublattice) representation.
In Sec.~\ref{sec:berry}, we formulate the $U(1)$ twist and define the many-body Berry phase,
including a discretized Wilson-loop expression for numerical evaluation.
In Sec.~\ref{Sec:NumRes}, we describe the DMRG setup and present numerical results for the
excitation gap, the quantum-metric estimator extracted from neighbor overlaps, and the
comparison between the many-body and density-matching KS descriptions.
Finally, we summarize the results and discuss implications for density-constrained KS
descriptions of geometric phases.

%% file: sections/02_model.tex
\section{The model}
\label{sec:model}
We consider a dimerized Fermi--Hubbard chain (SSH--Hubbard model) with an even number of sites $L=2N$ with an onsite repulsion $U \ge 0$.
The Hamiltonian reads
\begin{equation}
\label{eq:SSHHub_no_sublattice_i}
\hat H
=
-\sum_{i=1}^{2N}\sum_{\sigma}
\Bigl[
t_i\, c^{\dagger}_{i\sigma}c_{(i+1)\sigma}
+ \mathrm{h.c.}
\Bigr]
+U\sum_{i=1}^{2N} n_{i\uparrow}n_{i\downarrow},
\end{equation}
where
$c_{i\sigma}$ and $c_{i\sigma}^\dagger$ denote
electron annihilation and creation operators on site $i$ with spin $\sigma$, respectively.
These operators satisfy the following anti-commutation relations,
\begin{align}
\{ c_{i\sigma}, c_{j\sigma'}^\dagger \}
&= \delta_{ij}\delta_{\sigma\sigma'}, \\
\{ c_{i\sigma}, c_{j\sigma'} \}
&= \{ c_{i\sigma}^\dagger, c_{j\sigma'}^\dagger \}
= 0.
\end{align}
Number operator is defined as
\begin{equation}
n_{i\sigma} = c_{i\sigma}^\dagger c_{i\sigma}.
\end{equation}
The vacuum of the model $|0\rangle$ is defined by
\begin{equation}
c_{i\sigma}\,|0\rangle = 0
\qquad (\forall\, i,\sigma).
\end{equation}

We impose periodic boundary conditions (PBC),
\begin{equation}
c_{(2N+1)\sigma}\equiv c_{1\sigma},
\end{equation}
and introduce dimerization through the staggered hopping amplitudes,
\begin{equation}
t_i=t\left\{1+(-1)^i\delta\right\},
\label{eq:Staggered_Hopping}
\end{equation}
with a constant dimensionless parameter $\delta>0$.
To keep the bulk excitation gap well defined along the twist cycle, we work with PBC throughout.
The issue with open boundary conditions (OBC) is discussed in Sec.~\ref{Sec:NumRes}.

We focus on half filling, i.e., the total number of electrons is fixed to $N_e=2N$ (unless stated otherwise).
At half filling and in the spin-unpolarized sector, $\langle n_i\rangle\simeq 1$ and hence $\langle n_{i\sigma}\rangle\simeq 1/2$.

\subsection{Sublattice representation}
To make the dimerized structure explicit (and for later use in the KS construction), we introduce a unit-cell index and sublattice labels (A,B).
Namely,
\begin{equation}
c_{i,A\sigma} \equiv c_{(2i-1)\sigma}, \qquad
c_{i,B\sigma} \equiv c_{(2i)\sigma}
\end{equation}
so that Eq.~(\ref{eq:SSHHub_no_sublattice_i}) can be rewritten as
\begin{widetext}
\begin{equation}
\label{eq:SSH_Hubbard_sublattice_ti}
\hat{H}
=
-t\sum_{i=1}^{N}\sum_{\sigma}
\left[
\left(1-\delta\right)\,c_{i,A\sigma}^{\dagger}c_{i,B\sigma}
+
\left(1+\delta\right)\,c_{i,B\sigma}^{\dagger}c_{(i+1),A\sigma}
+\mathrm{h.c.}
\right]
+
U\sum_{i=1}^{N}\sum_{\alpha=A,B}
n_{i,\alpha\uparrow}n_{i,\alpha\downarrow},
\end{equation}
\end{widetext}
where the index $i$ now labels unit cells (each consisting of two sites), and the sum runs over the number of cells $N$.
In this representation, PBC reads,
\begin{equation}
c_{(N+1),\alpha\sigma} \equiv c_{1,\alpha\sigma}
\qquad (\alpha = A,B).
\end{equation}
In deriving Eq.~(\ref{eq:SSH_Hubbard_sublattice_ti}) we have used the definition of the staggered hopping amplitudes in Eq.~(\ref{eq:Staggered_Hopping}).
The number operator is correspondingly written with the cell and sublattice labels as
$n_{i,\alpha\sigma}\equiv c_{i,\alpha\sigma}^{\dagger}c_{i,\alpha\sigma}$.

We denote the eigenvalues and eigenstates of $\hat H$ by $E_n(U)$ and $\ket{\Psi_n(U)}$, respectively, satisfying
\begin{equation}
\label{eq:eigen}
\hat H \ket{\Psi_n(U)} = E_n(U)\ket{\Psi_n(U)}.
\end{equation}
Analytical results are available only in special limits, such as $U=0$ (spinful SSH model; Appendix~\ref{app:bloch}), $U\gg t$ (effective Heisenberg model), and $\delta=0$ (uniform one-dimensional Hubbard model solvable by the Bethe ansatz).
In this work, the eigenvalue problem is solved numerically using DMRG, as described in Sec.~\ref{Sec:NumRes}.

\section{$U(1)$ gauge twist (flux insertion) and many-body Berry phase}
\label{sec:berry}
In this section we describe how to define the many-body Berry phase from a $\theta$-parametrized family of ground states of $\hat H(\theta)$.
For the Berry phase to be well defined, the ground state must remain nondegenerate and separated from excited states by a finite excitation gap along the entire twist cycle $\theta\in[0,2\pi]$.
This condition is verified directly by numerical calculations using DMRG in Section ~\ref{Sec:NumRes}.

In the uniform-gauge convention, the Hamiltonian threaded by a
U(1) gauge flux $\theta$ is written in the sublattice representation as
\begin{widetext}
\begin{align}
\label{eq:local_gauge}
\hat H(\theta)
=
-t
\sum_{\sigma}
\Big[
\sum_{j=1}^{N}
\left\{
\left(
1-\delta
\right)
e^{-i\frac{\theta}{2N}}
c^\dagger_{j,A\sigma} c_{j,B\sigma}
+
\left(
1+\delta
\right)
e^{-i\frac{\theta}{2N}}
c^\dagger_{j,B\sigma} c_{(j+1),A\sigma}
\right\}
+\mathrm{h.c.}
\Big]
+
U\sum_{i=1}^{N}\sum_{\alpha=A,B}
n_{i,\alpha\uparrow} n_{i,\alpha\downarrow}.
\end{align}
\end{widetext}
The uniform factor $e^{-i\theta/(2N)}$ on each link is equivalent
to introducing the corresponding Peierls phase into the link variables.

Under PBC, the total twist along the ring cannot be gauged away by a single-valued transformation.
This is the origin of the Berry phase associated with the $\theta$ cycle~\cite{Berry1984,Zak1989}.

To make this explicit, we perform a  local transformation
\begin{equation}
c_{j\sigma} \to\, e^{i \theta \frac{j}{2N}} c_{j\sigma}.
\end{equation}
 This gauge transformation removes the uniform bulk phase and shifts the twist to the boundary link~\cite{Zawadzki2017}.
 For the boundary hopping, we obtain
\begin{equation}
c_{N,B\sigma}^{\dagger}c_{1,A\sigma}
\to
e^{-i\theta\frac{2N}{2N}}
e^{i\theta\frac{1}{2N}}
c_{N,B\sigma}^{\dagger}c_{1,A\sigma}
=
e^{-i\theta}
e^{i\frac{\theta}{2N}}
c_{N,B\sigma}^{\dagger}c_{1,A\sigma}.
\end{equation}
Thus the bulk hoppings become phase free, while the boundary hopping carries the entire twist.

Accordingly, Eq.~\eqref{eq:local_gauge} can be rewritten in the boundary-twist form,
\begin{widetext}
\begin{align}
\label{eq:boundary_gauge}
\hat H(\theta)
=
-t\sum_{\sigma}
\Big[
\sum_{i=1}^{N-1}
\Big\{
(1-\delta)\,
c^\dagger_{i,A\sigma} c_{i,B\sigma}
+
(1+\delta)\,
c^\dagger_{i,B\sigma} c_{(i+1),A\sigma}
\Big\}
\nonumber\\
\qquad\qquad
+
(1+\delta)\,e^{-i\theta}\,
c^\dagger_{N,B\sigma} c_{1,A\sigma}
+ \mathrm{h.c.}
\Big]
+
U\sum_{i=1}^{N}\sum_{\alpha=A,B}
n_{i,\alpha\uparrow} n_{i,\alpha\downarrow}.
\end{align}
\end{widetext}
This form is equivalent to imposing a twisted boundary condition with twist angle $\theta$~\cite{LiebSchultzMattis1961,NiuThoulessWu1985,RestaSorella1999}.
Only the total phase accumulated around the ring is physically meaningful: Eqs.~\eqref{eq:local_gauge} and \eqref{eq:boundary_gauge} are related by a unitary gauge transformation and hence describe the same physics.

The uniform-phase form in Eq.~\eqref{eq:local_gauge} is convenient for numerical calculations, and we use it in the following sections unless otherwise noted.

\subsection{Many-body Berry phase}
\label{sec:mb_berry}

The ground state of $\hat H(\theta)$ satisfies
\begin{equation}
\label{eq:eigen_theta}
\hat H(\theta)\ket{\Psi_0(\theta)} = E_0(\theta)\ket{\Psi_0(\theta)}.
\end{equation}
Treating $\theta$ as an adiabatic parameter, we consider the cycle $\theta:0\to 2\pi$, along which the ground state acquires a geometric (Berry) phase
\begin{equation}
\label{eq:zak_mb}
\gamma
=
i\int_{0}^{2\pi}
\langle \Psi_0(\theta)\vert \partial_\theta \Psi_0(\theta)\rangle\, d\theta
\quad (\mathrm{mod}\ 2\pi).
\end{equation}


Introducing $\theta$ as an adiabatic parameter does not merely attach a phase to a fixed wave function.
Rather, it defines a continuous family of Hamiltonians $\hat H(\theta)$ through a $U(1)$ gauge twist (equivalently, a flux insertion), and the corresponding ground state $\ket{\Psi_0(\theta)}$ is followed along the $\theta$ cycle.
Conceptually, $\theta$ should be viewed as a parameter that labels a path in Hamiltonian space.

The corresponding ground state $\ket{\Psi_0(\theta)}$ is obtained by projecting to the ground-state manifold of $\hat H(\theta)$ at each $\theta$, so that the $\theta$ cycle defines an adiabatic walk on the resulting family of ground states. In the uniform-gauge convention, the total twist $\theta$ is distributed evenly over nearest-neighbor links, so that each hopping acquires a phase factor $e^{i\theta/L}$ (with $L$ the number of sites).
Accordingly, updating $\theta_j\to\theta_{j+1}$ changes each link phase by $\Delta\theta/L$, which is locally infinitesimal, in particular in the $L\to\infty$ limit.

As long as the excitation gap remains open, the ground state varies smoothly along the cycle.
On a discretized theta grid, namely $\{\theta_j\}$, the neighbor overlap (Wilson-loop)
\[
\mathcal{O}_j \equiv \langle \Psi_0(\theta_j)\vert\Psi_0(\theta_{j+1})\rangle
\]
quantifies the local deformation of the state along the $\theta$ direction.
In particular, deviations of $|\mathcal{O}_j|$ from unity indicate a nontrivial flux sensitivity of the wave function.
On the other hand, the phase of the product
\[
\prod_j \mathcal{O}_j
\]
yields the Berry phase associated with one cycle $\theta:0\to 2\pi$, which is interpreted as the many-body polarization.
Accordingly, the adiabatic $\theta$ cycle is consistent with the charge-pump picture of polarization.

For the Berry phase to be well defined, the ground state must remain nondegenerate and separated from excited states by a finite energy gap throughout the cycle $\theta\in[0,2\pi]$~\cite{Gonze1997}.
From the viewpoint of analytic continuation in $\theta$, this gap condition keeps the Riemann sheet associated with the ground-state energy separated from those of excited states, i.e., the corresponding analytic branch does not intersect excited-state branches under a continuous variation of $\theta$.
This guarantees that the ground state can be followed continuously around the $\theta$ cycle (up to an overall phase); that is why the Berry phase is well defined.

This condition is manifest in two limiting cases.
At $U=0$ and $\delta\neq 0$, the system is a band insulator at half filling with a unique many-body ground state separated from excited states by the single-particle gap.
In the opposite limit $U\to\infty$, charge fluctuations are suppressed and the low-energy sector is described by an alternating Heisenberg chain, which remains gapped and nondegenerate.

Since both limits belong to gapped phases preserving the same symmetries, the ground-state branch is expected to remain isolated along the entire $\theta$ cycle in the absence of a phase transition.
In the present work, this expectation is verified numerically by monitoring the excitation gap as a function of $\theta$ and $U$ (see Sec.~\ref{Sec:NumRes}).

\subsection{Discretized definition for numerical calculations}
For numerical evaluation, the Berry phase is computed by discretizing $\theta$ into $N_\theta$ grids $\{\theta_j\}$ with $\theta_{N_\theta+1}=\theta_1$ and evaluating~\cite{SouzaWilkensMartin2000}
\begin{equation}
\gamma
=
-\mathrm{Im}
\log
\prod_{j}
\mathcal{O}_j.
\label{eq:Berry_phase_diff}
\end{equation}

The equivalence to Eq.~\eqref{eq:zak_mb} follows from a small-$\Delta\theta$ expansion.
For neighboring points $\theta_j$ and $\theta_{j+1}=\theta_j+\Delta\theta$, one finds
\[
\log \mathcal{O}_j
=
\Delta\theta\,\langle \Psi_0(\theta_j) | \partial_\theta \Psi_0(\theta_j) \rangle
+O(\Delta\theta^2).
\]
Summing over $j$ gives
\[
\sum_{j=0}^{N_\theta-1}\log \mathcal{O}_j
=
\Delta\theta \sum_{j=0}^{N_\theta-1}\langle \Psi_0(\theta_j) | \partial_\theta \Psi_0(\theta_j) \rangle
+O(\Delta\theta).
\]
Using $\sum_j \log \mathcal{O}_j=\log\prod_j \mathcal{O}_j$, the left-hand side becomes $\log\prod_{j=0}^{N_\theta-1}\mathcal{O}_j$.
The right-hand side is a Riemann sum and converges to the continuum integral in Eq.~\eqref{eq:zak_mb} as $\Delta\theta\to0$.

\subsection{Kohn--Sham construction}
The aim of this study is to assess, in a controlled setting, 
how strongly the many-body Berry phase is constrained 
when one restricts the description to the ground-state density 
as the sole basic variable. 
To this end, we fix the ground-state density obtained numerically 
for the interacting many-body system,
\[
n_i^{\mathrm{MB}}(\theta)
=
\langle \Psi_0(\theta)|
\hat n_i
|\Psi_0(\theta)\rangle,
\]
and construct a noninteracting Kohn--Sham (KS) effective system 
whose ground state reproduces the same density for all $\theta$.

In a general lattice model,
one introduces a noninteracting Hamiltonian of the form
\begin{equation}
\hat H_{\mathrm{KS}}
=
\hat H_{\mathrm{hop}}
+
\sum_{i,\sigma}
v_i \hat n_{i\sigma},
\end{equation}
and determines the set of local potentials $\{v_i\}$ such that
\begin{equation}
n_i^{\mathrm{KS}}(\theta)
=
n_i^{\mathrm{MB}}(\theta)
\quad
\forall i.
\end{equation}
Here, $\hat H_{\mathrm hop}$ and $n_i^{\mathrm{KS}}(\theta)$ denote the hopping term in Eq.~(\ref{eq:SSHHub_no_sublattice_i}) and Kohn--Sham density respectively.
Without additional symmetry constraints, determining $\{v_i\}$ constitutes
a multidimensional density-inversion problem, which is typically treated
numerically as a variational optimization over the potentials $\{v_i\}$.

\subsubsection{Structure of the KS potentials in the present model}
In the present SSH--Hubbard chain, however, the many-body density inherits the spatial
symmetries of the Hamiltonian under periodic boundary conditions at half filling.
As a result, the admissible KS potentials that can reproduce $n_i^{\mathrm{MB}}(\theta)$
are strongly constrained, and the inversion problem simplifies accordingly.
In particular, the dimerized hopping implies a two-site translational symmetry, which
reduces the density profile to two inequivalent sublattice values.

This can be shown as follows.
First, we define the two-site translation operator $\hat T_2$ by
\[
\hat T_2 \hat c_{j\sigma} \hat T_2^{-1}
=
\hat c_{(j+2)\sigma},
\]
one has
\[
[\hat H,\hat T_2]=0.
\]
Assuming the ground state to be non-degenerate, Wigner's theorem implies
\[
\hat T_2|\Psi_0\rangle
=
e^{iK}|\Psi_0\rangle,
\]
with a phase $K$.
Since the local density operator satisfies
\[
\hat T_2 \hat n_{j\sigma} \hat T_2^{-1}
=
\hat n_{(j+2)\sigma},
\]
it follows that
\[
\langle \hat n_{(j+2)\sigma}\rangle
=
\langle \hat n_{j\sigma}\rangle.
\]
In this model we work in the spin-unpolarized sector, i.e.,
$\langle \hat n_{j\uparrow}\rangle=\langle \hat n_{j\downarrow}\rangle$,
so that $\langle \hat n_j\rangle=\sum_{\sigma}\langle \hat n_{j\sigma}\rangle$.
Therefore,
\[
\langle \hat n_{j+2}\rangle=\langle \hat n_j\rangle,
\]
and the ground-state density is fully characterized by two inequivalent values
(even/odd sites, or equivalently $A/B$ sublattices),
\begin{equation}
n^{\mathrm{KS}}_i
=
\left(
n^{\mathrm{KS}}_{\mathrm{even}},
\,n^{\mathrm{KS}}_{\mathrm{odd}}
\right).
\end{equation}

In the uniform-gauge convention, Eq.~(\ref{eq:local_gauge}), the phase factors are
distributed uniformly and the dimerized hoppings satisfy $t_{j+2}=t_j$; hence the two-site
translation symmetry is manifest:
\[
[\hat H(\theta),\hat T_2]=0
\quad
(\forall\,\theta).
\]

At half filling,
\[
\frac{1}{L}\sum_j \langle \hat n_j\rangle=1,
\]
so that the two densities $n_A,n_B$ satisfy
\[
n_A+n_B=2.
\]
Hence, the independent degree of freedom reduces to
\[
\delta n=n_A-n_B.
\]

Under these symmetry constraints and imposing inversion symmetry, the most general one-body potential takes the form
\[
v_i=
\begin{cases}
+\Delta_{\mathrm{KS}} & (i\in A),\\
-\Delta_{\mathrm{KS}} & (i\in B),
\end{cases}
\]
and the density inversion problem reduces from a multidimensional optimization to a single-variable root search solving
\[
\delta n_{\mathrm{KS}}(\Delta_{\mathrm{KS}})
=
\delta n_{\mathrm{MB}}.
\]
Thus, while the KS construction is formally a general density-inversion problem,
in the present model it is rigorously reduced to a one-parameter problem.

In momentum space, the Bloch Hamiltonian reads
\[
H_{\mathrm{KS}}(k)
=
\begin{pmatrix}
\Delta_{\mathrm{KS}} & t_1+t_2 e^{-ik}\\
t_1+t_2 e^{ik} & -\Delta_{\mathrm{KS}}
\end{pmatrix}
\]
with eigenvalues
\[
E_\pm(k)
=
\pm
\sqrt{
\Delta_{\mathrm{KS}}^2
+
t_1^2+t_2^2+2t_1 t_2 \cos k
}.
\]

Therefore, the density inversion problem originating from the interacting SSH--Hubbard model reduces, without invoking mean-field approximations or perturbative expansions, to a one-dimensional single-particle KS equation with a period-two potential.


%% file: sections/03_method.tex

%% file: sections/04_results.tex
\section{Numerical results}
\label{Sec:NumRes}
\subsection{DMRG setup}
We solve the many-body eigenvalue problem for $\hat H(\theta)$ using the density-matrix
renormalization group (DMRG) method
\cite{White1992,White1993,Schollwoeck2011}.
Our implementation is based on the TenPy library
\cite{HauschildPollmann2018,TenPyDocs,TenPyGitHub}.

We fix the hopping amplitude to $t=1$ as the energy unit and work with the dimerization
$\delta=0.6$ (Sec.~\ref{sec:model}).
All results are presented in terms of dimensionless ratios such as $U/t$.
In the noninteracting limit ($U=0$), $\delta$ controls the single-particle gap; the present
choice yields a well-developed insulating regime and stabilizes the evaluation of the
many-body Berry phase along the twist cycle.
The representative values of $U$ used in this study are summarized in Appendix~\ref{app:strength_U}.

\subsubsection*{Data Availability}
All numerical data underlying the figures are available from the author upon reasonable request.
\subsubsection{Benchmark against exact diagonalization}

For small system sizes, we benchmark the DMRG results against exact 
diagonalization (ED). 
At half filling, ED becomes impractical beyond
$L = 2N \gtrsim 20$ due to the exponential growth of
the Hilbert-space dimension.

For $L = 2N = 12$, where ED is feasible,
we find that the ground-state energies obtained by DMRG agree with ED under periodic
boundary conditions (PBC) within numerical precision, with deviations smaller than $10^{-13}$.
The same level of agreement is observed for $L=6,8,10$.

These comparisons validate the correctness of the hopping terms, 
interaction terms, and MPO construction.


\subsubsection{Choice of Boundary conditions}

For open boundary conditions (OBC), the excitation gap decreases rapidly with increasing
system size. This behavior is expected in the topological regime: edge-localized zero modes
emerge, leading to an exponentially small splitting between the ground state and the first
excited state. For the largest system size accessible by exact diagonalization, $L=12$,
the gap is of order $\Delta=E_1-E_0\sim 10^{-4}$, which is an indication of the edge mode.

In contrast, under periodic boundary conditions (PBC), boundary effects are removed and the
bulk gap remains finite. This ensures a nondegenerate ground state separated by a finite gap,
which is required for a well-defined many-body Berry phase. We therefore restrict all
subsequent calculations to PBC.

\subsubsection{Convergence and system-size dependence}

System-size dependence was examined for $L=8,12,20,30,40,50$.
For $L\ge 20$, physical observables show only minor variations and stable numerical
convergence is obtained.

For the representative setup $L=20$, the orthogonality condition between the ground and
first excited states satisfies $|\langle \psi_0|\psi_1\rangle|\lesssim 5.0\times 10^{-7}$ for all
$U$, and the energy consistency condition
\[
\left|
E^{\mathrm{eigen}}(\theta)
-
\langle \psi|\hat H(\theta)|\psi\rangle
\right|
\le 10^{-10}
\]
is fulfilled within machine precision.

For PBC with an odd number of unit cells ($N$ odd), $|\langle \psi_0|\psi_1\rangle|$ increases
compared to even sizes, reflecting finite-size effects associated with the dimerization
pattern. We therefore adopt $L=20$ (even $N$) as the representative system size.

The DMRG setup is summarized in Table.~\ref{tab:dmrg_params}.

\begin{table*}[t]
\caption{DMRG simulation parameters in TeNPy for the ground state (GS) and the first excited state (ES). The ES was further refined by additional sweeps to ensure orthogonality and convergence.}
\label{tab:dmrg_params}
\begin{ruledtabular}
\begin{tabular}{ll}
Parameter & Value \\ \hline
Maximum bond dimension $m$ & 300 (GS) and 500 (ES)\\
Truncation error & $\le 10^{-8}$ (GS), $10^{-12}$ (ES), and $10^{-14}$ (ES refinement) \\
Number of sweeps & 60 (GS), 90 (ES), and 60 (ES refinement)\\
Energy convergence threshold & $10^{-8}$ for both GS and ES\\
Twist implementation and boundary condition & uniform gauge and periodic boundary condition (PBC)\\
\end{tabular}
\end{ruledtabular}
\end{table*}


\subsection{Many-body Wave function from DMRG and Phase evaluation}
After fixing the representative setup, we compute the many-body ground state
$\ket{\Psi_0(\theta,U)}$ on a discretized twist grid $\theta_j=2\pi j/N_\theta$ for each $U$
using DMRG.

\subsubsection{Excitation energy gap}
For the Berry phase to be well defined, the excitation gap
\[
\Delta_{\mathrm{MB}}(\theta,U)=E_1(\theta,U)-E_0(\theta,U)
\]
must remain finite along the entire twist cycle.

In the following, we track the $\theta$-dependent ground state $|\Psi_0(\theta,U)\rangle$
on a discretized twist grid and extract the excitation gap and the Berry phase as functions
of $U$.

\begin{figure}[t]
  \centering
  \includegraphics[width=1.0\columnwidth]{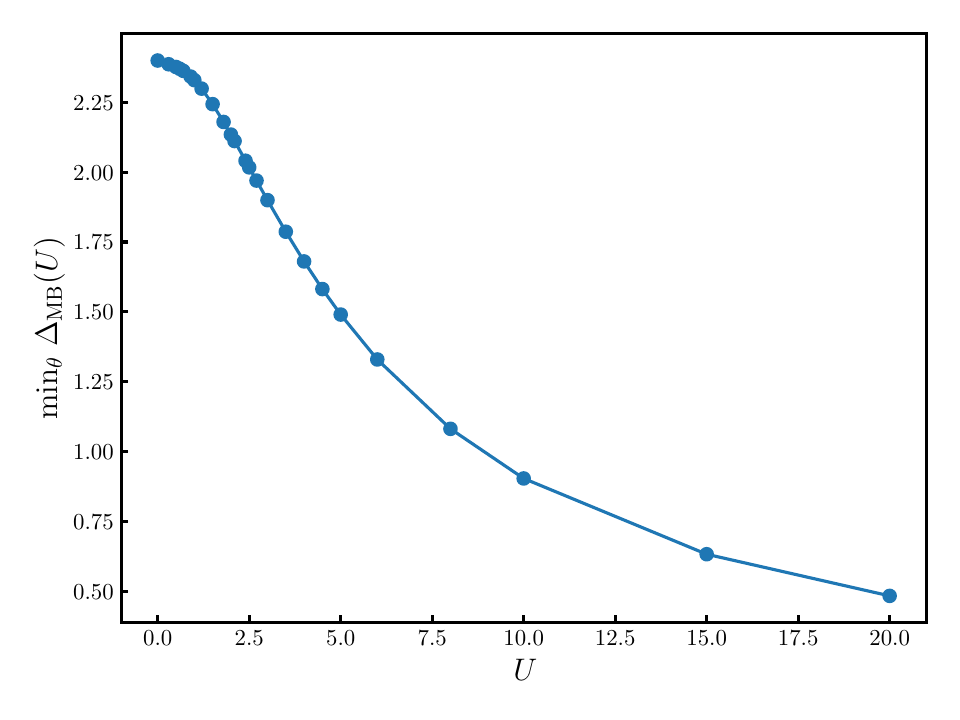}
\caption{Minimum excitation gap along the twist cycle.
For each $U$, we evaluate $\Delta_{\mathrm{MB}}^{\min}(U)\equiv \min_{\theta_j}\Delta_{\mathrm{MB}}(\theta_j,U)$ on the discretized twist grid and plot it as a function of $U$.}
  \label{fig:gapU}
\end{figure}

Since the excitation gap $\Delta_{\mathrm{MB}}(\theta,U)=E_1(\theta,U)-E_0(\theta,U)$
depends on both $U$ and $\theta$, the Berry phase is well defined only if the gap does not
close for any $\theta$ along the twist cycle. We therefore evaluate, for each fixed $U$,
the minimum gap on the discretized twist grid,
\[
\Delta_{\mathrm{MB}}^{\min}(U)\equiv \min_{\theta_j}\Delta_{\mathrm{MB}}(\theta_j,U),
\]
and plot $\Delta_{\mathrm{MB}}^{\min}(U)$ as a function of $U$ in Fig.~\ref{fig:gapU}.
The finiteness of $\Delta_{\mathrm{MB}}^{\min}(U)$ confirms that the gap remains open for all
$\theta$ throughout the $U$ range studied.

\setlength{\textfloatsep}{10pt}      
\setlength{\abovecaptionskip}{4pt}   
\setlength{\belowcaptionskip}{10pt}   

\subsubsection{Quantum metric from neighbor overlaps}
Once the many-body ground state is obtained, we track its evolution along the twist cycle
via the neighbor overlaps
\[
\mathcal{O}_j(U)\equiv
\braket{\Psi_0(\theta_j,U)|\Psi_0(\theta_{j+1},U)}.
\]
These overlaps enter the Wilson loop evaluation of the Berry phase and provide the basic
input for local geometric diagnostics.

While $\mathcal{O}_j(U)$ directly probes the response to the gauge twist $\theta$, its magnitude
approaches unity as $\Delta\theta\to 0$, $|\mathcal{O}_j(U)|\to 1$, with deviations scaling as
$O\!\left((\Delta\theta)^2\right)$. Hence the raw overlap becomes progressively insensitive as
the twist grid is refined.

To quantify the local geometric response along the twist direction, we define a link-resolved
estimator of the quantum metric~\cite{Provost1980,Wang2023QuantumMetric},
\begin{equation}
g^{\rm (est)}_{\theta\theta}(\theta_j,U)
\equiv
-\frac{2}{(\Delta\theta)^2}\log|\mathcal O_j(U)|,
\qquad
\Delta\theta=\frac{2\pi}{N_\theta}.
\label{eq:g_est}
\end{equation}
In the continuum limit $\Delta\theta\to 0$, $g^{\rm (est)}_{\theta\theta}(\theta_j,U)$ reproduces the
quantum metric.


To eliminate branch-matching artifacts associated with the cycle-closure (wrap) link
$\theta_{N_\theta-1}\to\theta_0$, we define the all-link mean
\[
\langle g^{(\mathrm{est})}_{\theta\theta}\rangle_{\rm all}(U)
\equiv
\frac{1}{N_\theta}\sum_{j=0}^{N_\theta-1} g^{(\mathrm{est})}_{\theta\theta}(\theta_j,U),
\]
and the unwrap-link mean by excluding the unique wrap link (i.e., the link with $\theta_{j+1}<\theta_j$),
\[
\langle g^{(\mathrm{est})}_{\theta\theta}\rangle_{\rm unwrap}(U)
\equiv
\frac{1}{N_\theta-1}\sum_{j=0}^{N_\theta-2} g^{(\mathrm{est})}_{\theta\theta}(\theta_j,U).
\]
All local geometric diagnostics are evaluated using $\langle g^{(\mathrm{est})}_{\theta\theta}\rangle_{\rm unwrap}(U)$, while the Berry phase is evaluated from the full Wilson loop as in Eq.~(\ref{eq:Berry_phase_diff}).

\begin{figure}[t]
  \centering
  \includegraphics[width=\columnwidth]{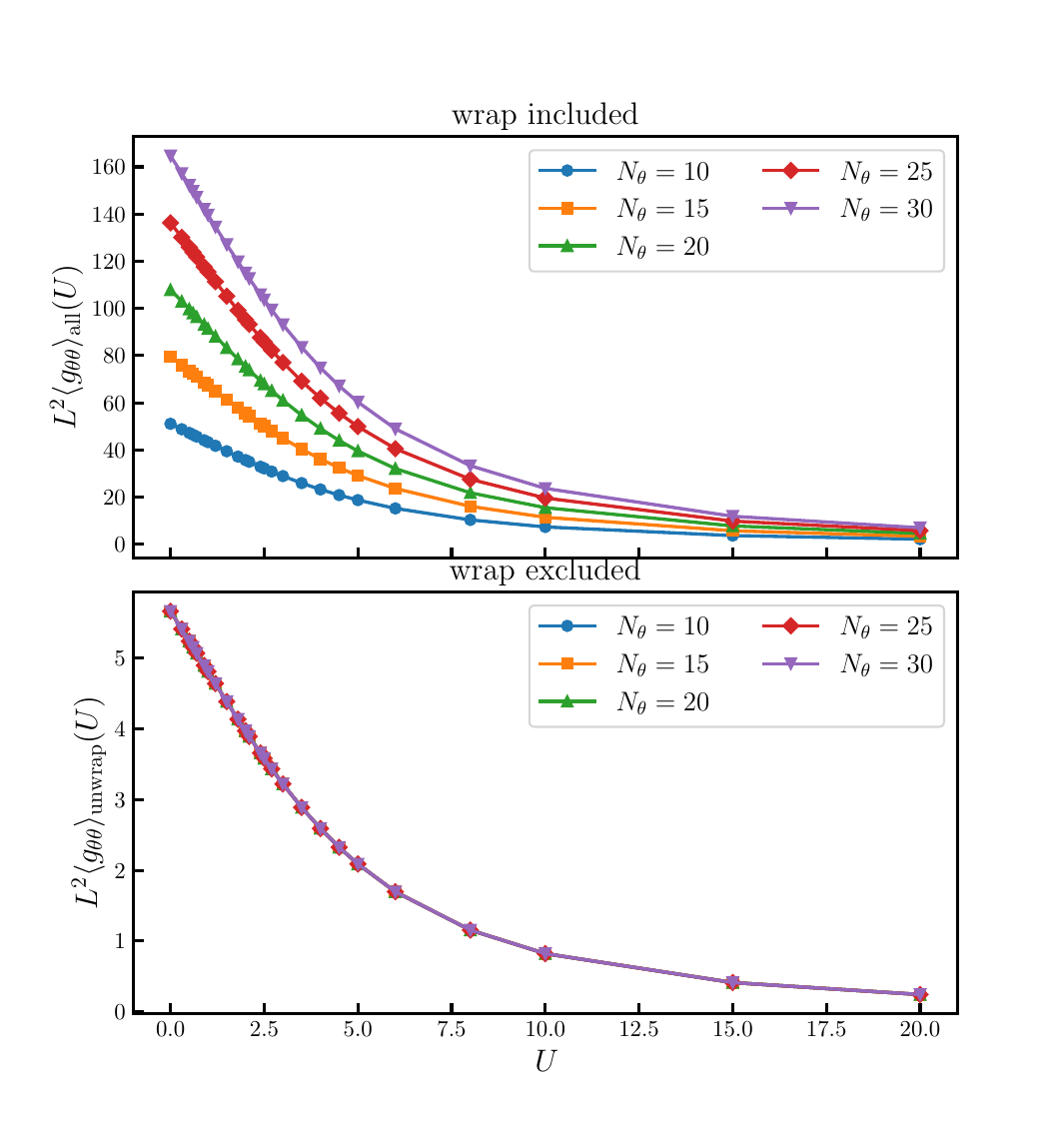}
\caption{Quantum metric $L^2\langle g_{\theta\theta}\rangle(U)$ as function of $U$. Here $\langle\cdots\rangle_{\rm unwrap}$ denotes the mean over the discretized twist links excluding the wrap-around link $\theta_{N_\theta-1}\to\theta_0$.
}
  \label{fig:geometry_conv}
\end{figure}

As shown in Fig.~\ref{fig:geometry_conv}, $L^2\langle g_{\theta\theta}\rangle_{\rm unwrap}(U)$ converges smoothly to a single curve
as $N_\theta$ increases and becomes numerically indistinguishable for $N_\theta\ge 20$.
In contrast, $L^2\langle g_{\theta\theta}\rangle_{\rm all}(U)$ exhibits a nonuniform dependence on $N_\theta$.
This behavior is not a physical effect but originates from the
cycle-closure link $\theta_{N_\theta-1}\to\theta_0$,
where the gauge/branch matching of the ground-state manifold is most exposed.
The wrap link is therefore intrinsically sensitive to discretization artifacts,
whereas the unwrap links probe the local deformation of the wave function
along the twist trajectory.

Accordingly, all local geometric diagnostics are evaluated on unwrap links only.
By contrast, the many-body Berry phase is computed from the Wilson loop,
i.e., the product of overlaps over \emph{all} links along the closed cycle,
as required by its definition.
The $\theta$ grid points are set to $N_\theta=20$ from now on.

\setlength{\textfloatsep}{10pt}      
\setlength{\abovecaptionskip}{4pt}   
\setlength{\belowcaptionskip}{10pt}   

Having established that $\langle g_{\theta\theta}\rangle_{\rm unwrap}(U)$ provides a stable
local diagnostic free from the cycle-closure artifact, we now discuss the expected
interaction dependence of the quantum metric. 
The twist $\theta$ enters $\hat H(\theta)$ only through the hopping term, while the interaction term is $\theta$-independent.
In the strong-coupling regime $U\gg t$, the twist response is controlled by virtual charge processes.
Standard perturbation theory in the parameter $\theta$ yields the spectral representation of the geometric response,
\[
g_{\theta\theta}(\theta)
=
\sum_{n\neq 0}
\frac{
\left|\bra{\Psi_n(\theta)}\,\partial_\theta \hat H(\theta)\,\ket{\Psi_0(\theta)}\right|^2
}{
\left(E_n(\theta)-E_0(\theta)\right)^2
}.
\]
From the model Hamiltonian we find $\partial_\theta \hat H=O(t)$, while the relevant intermediate states involve charge excitations with an energy cost $E_n(\theta)-E_0(\theta)=O(U)$, this immediately implies the parametric estimate
\[
g_{\theta\theta}\sim \frac{t^2}{U^2},
\]
up to a dimensionless prefactor set by the dimerization pattern and system size.
This explains why, for $U\gtrsim 5$, the $U$ dependence of $L^2\langle g_{\theta\theta}\rangle_{\rm unwrap}(U)$ resembles that of the excitation gap on the strong-coupling side, apart from an overall scale.

Consistent with this expectation, for $U\gtrsim 5$ the quantum metric indicator
$\langle g_{\theta\theta}\rangle_{\rm unwrap}(U)$ decreases monotonically and tracks the suppression of low-energy fluctuations with increasing $U$
(see the lower panel of Fig.~\ref{fig:geometry_conv}).
The apparent similarity between $\langle g_{\theta\theta}\rangle_{\rm unwrap}(U)$ and the excitation gap in the strong-coupling regime reflects the dynamical suppression of charge fluctuations:
at large $U$, $\Delta_{\rm neu}\sim t^2/U$ whereas $g_{\theta\theta}^{(\log)}\sim (t/U)^2$.

For small $U$, the ground state is well approximated by a Slater determinant of
single-particle Bloch states, i.e., it is close to the noninteracting SSH limit.
In this regime the $U(1)$ twist effectively shifts the crystal momentum as
$k\to k+\theta/L$, so the $\theta$ dependence of the many-body state is inherited directly
from the single-particle band structure.
For the SSH model the dispersion reads
\[
\varepsilon_\pm(k)=\pm\bigl|t_1+t_2 e^{ik}\bigr|
=\pm\sqrt{t_1^2+t_2^2+2t_1t_2\cos k},
\]
and the resulting momentum sensitivity under $k\to k+\theta/L$ leads to an enhanced
$\theta$ dependence of the weak-coupling wave function and the associated geometric
diagnostics.

\begin{figure}[t]
  \centering
  \includegraphics[width=1.0\columnwidth]{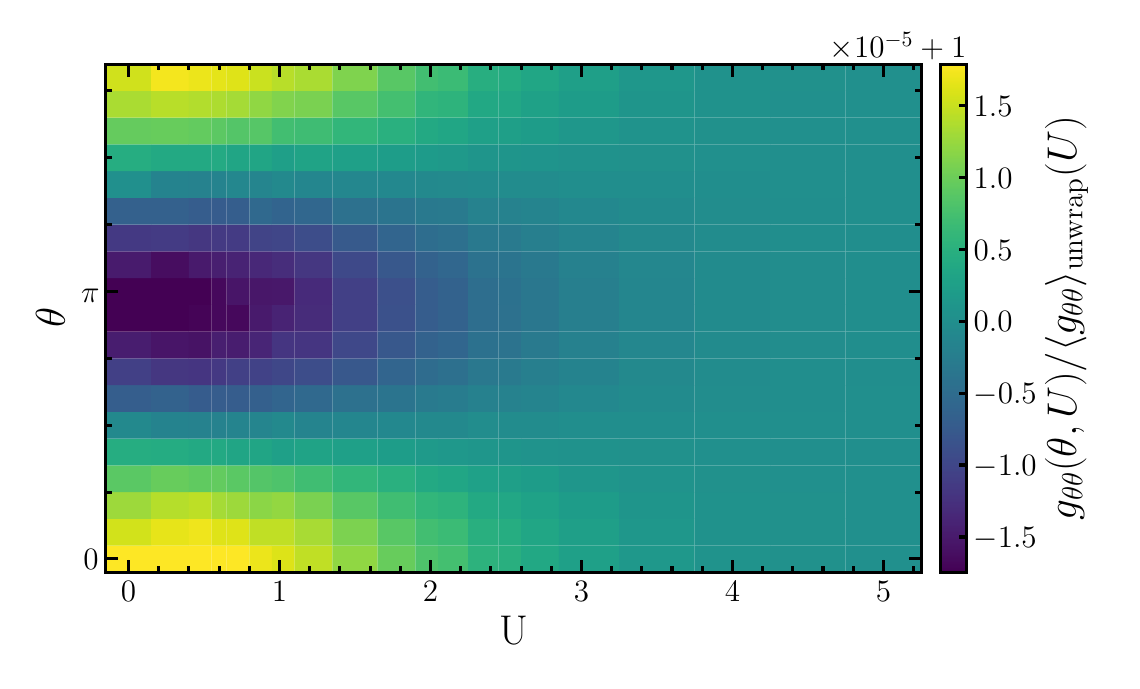}
  \caption{
Heatmap of the geometric response $g_{\theta\theta}(\theta,U)$, normalized by the unwrap average
$\langle g_{\theta\theta}\rangle_{\rm unwrap}(U)$ at each fixed $U$.
The normalization is introduced solely for visualization purposes, to improve the visibility of the $\theta$ dependence.
We restrict to the weak-coupling regime $U\le 5$, since at larger $U$ the $\theta$ variation becomes too small to be visually resolved on the same scale.
  }
  \label{fig:geom_heatmap}
\end{figure}

The heat map in Fig.~\ref{fig:geom_heatmap} shows the $\theta$-resolved geometric indicator
$g_{\theta\theta}(\theta,U)/\langle g_{\theta\theta}\rangle_{\rm unwrap}(U)$ as a function of $\theta$ and $U$.
In the weak-coupling regime, the geometric response is enhanced, consistent with the stronger sensitivity of the many-body ground state to flux insertion.
Upon increasing $U$, the geometric response is systematically suppressed.
In the strong-coupling regime, charge fluctuations are reduced and the current $\partial_\theta \hat H$ predominantly couples the ground state to charge excitations with an energy scale set by the charge gap; consequently $g_{\theta\theta}$ decreases with $U$ and tends toward zero in the large-$U$ limit.
The pattern is approximately symmetric about $\theta=\pi$, reflecting the $\theta\rightarrow 2\pi-\theta$ symmetry of the twisted Hamiltonian in the inversion-symmetric setting considered here.

\subsection{Solving Kohn--Sham equation}
In the present setup, the site densities $n_A$, $n_B$, and $n_i$ obtained from the many-body ground state show no resolvable dependence on $\theta$ or $U$ within our numerical accuracy: the residual variations are at the level of $\sim 10^{-9}$ (with a maximum deviation $\lesssim 10^{-8}$), and the densities remain pinned to $\langle \hat n_{i\sigma}\rangle \simeq 0.5$.
This indicates that the densities show no measurable dependence on $\theta$ and $U$.
Consequently, the density constraint provides no nontrivial information for determining a
symmetry-preserving KS potential beyond an irrelevant constant shift.
The natural choice for the KS potential is therefore
\[
\Delta_{\rm KS}=0,
\]
within numerical precision.

Therefore, in the present regime the density-constrained KS reference reduces to the SSH model
itself with $\Delta_{\rm KS}=0$, i.e., a purely quadratic free-fermion problem.
Accordingly, the KS ground state $|\Phi_{\rm KS}(\theta)\rangle$ is obtained by exact
diagonalization of the single-particle Hamiltonian, without invoking DMRG.
As a basic consistency check, the KS density computed from the occupied single-particle
orbitals reproduces the many-body density within numerical precision.
At $U=0$, where the KS reference coincides with the many-body model, the ground-state energies
also agree within machine precision.

Because the KS Hamiltonian is $U$ independent in the present setup, the resulting KS geometry is
likewise $U$ independent: the KS geometry is identical to that at $U=0$ for all $U$.

The interaction dependence of the geometric response is therefore not captured in the
density-constrained KS description in the present regime.
Nevertheless, the Berry phase can still agree with the many-body result throughout the $U$ range
studied because it is symmetry-quantized: the agreement is thus symmetry-enforced rather than
interaction-resolved.

\begin{figure}[t]
  \centering
  \includegraphics[width=1.0\columnwidth]{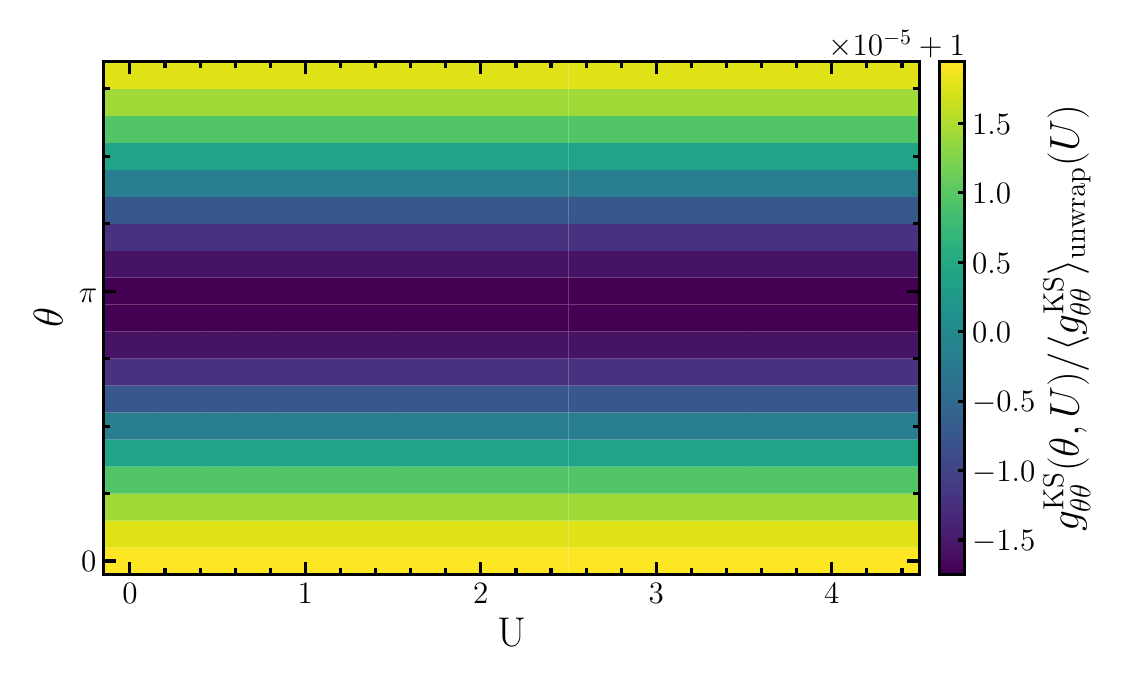}
  \caption{Heatmap of the KS geometric response $g^{\mathrm{KS}}_{\theta\theta}(\theta,U)$, normalized as in Fig.~\ref{fig:geom_heatmap}.
    KS data are shown for the baseline $U$ set listed in Appendix~\ref{app:strength_U}.
  }
  \label{fig:KS_geom_heatmap}
\end{figure}

To evaluate the Berry phase and geometric diagnostics within the KS effective system, we
insert the same twist $\theta$ into the KS Hamiltonian and compute the neighbor overlaps
\[
O^{\rm KS}_j \equiv \langle \Phi_{\rm KS}(\theta_j)|\Phi_{\rm KS}(\theta_{j+1})\rangle.
\]
From these overlaps we construct the KS geometric indicator
$g^{\rm KS}_{\theta\theta}(\theta_j,U)$ via Eq.~(\ref{eq:g_est}).
Fig.~\ref{fig:KS_geom_heatmap} shows the $(\theta,U)$ heatmap of the KS geometric response
in direct analogy with Fig.~\ref{fig:geom_heatmap}.
Since the KS Hamiltonian is $U$ independent in the present setup, the KS heatmap retains
essentially the same pattern for all $U$ and shows no suppression with increasing $U$,
in sharp contrast to the many-body result.

\begin{figure}[t]
    \centering
    \includegraphics[width=1.0\columnwidth]{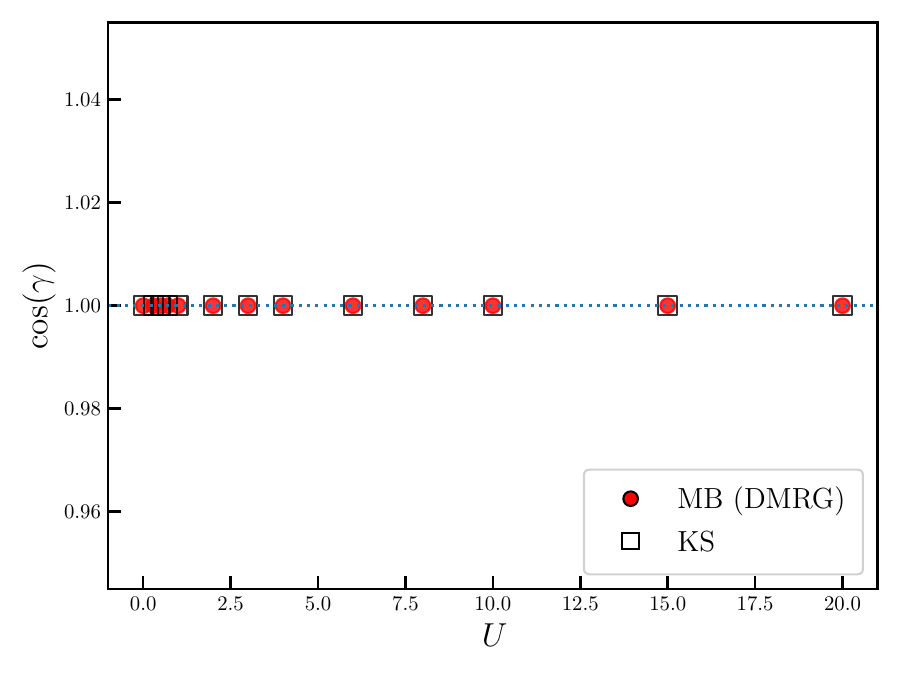}
\caption{
$\cos(\gamma)$ as a function of $U$ for the interacting many-body state (MB, DMRG) and the Kohn--Sham (KS) reference.
Filled circles denote MB results, while open squares denote KS results.
The horizontal line at $y=1$ serves as a visual guide.
Data are plotted for representative values of $U$ (Appendix~\ref{app:strength_U}).
}
    \label{fig:raw_gamma_over_2pi}
\end{figure}

The Berry phase is evaluated from the Wilson loop of the neighbor overlaps, $\mathcal{O}_j$ for the many-body state and $\mathcal{O}^{\mathrm{KS}}_j$ for the KS effective system.
Within our numerical resolution, we find $\gamma=0$ (mod $2\pi$) for both the KS and many-body states.
Fig.~\ref{fig:raw_gamma_over_2pi} shows $\cos\gamma$ extracted from the Wilson loop, confirming the agreement between the KS and many-body evaluations.

%% file: sections/05_discussion.tex
\clearpage
\section{Discussions and future prospects}

In this work, our model is constructed such that
a well-defined many-body Berry phase exists
under the insertion of a $U(1)$ gauge flux.
Under this premise, the ground state
$\ket{\Psi_0(\theta,U)}$
forms a continuous family as the flux parameter $\theta$
is swept over a full cycle.
That is, the interacting ground state traces out
a geometric ``trajectory'' in Hilbert space.

The finiteness of the quantum metric $g_{\theta\theta}$
confirms that this evolution is not merely a global phase rotation.
Therefore, the system exhibits a genuine geometric response
of the many-body wave function under flux insertion.

At the same time, the local density $n_i(\theta,U)$
and the sublattice density imbalance $n_A-n_B$
remain independent of both $\theta$ and $U$
within numerical accuracy.
This establishes a separation:
the density remains fixed,
while the wave function evolves geometrically.
Such a separation demonstrates that diagonal one-body observables,
including the density,
do not in general retain the geometric information
encoded in the many-body wave function,
such as the Berry connection and the quantum metric.

One may then ask:
if the density does not encode the geometric information of the wave function,
why do the Berry phases coincide?

For one-dimensional gapped systems with inversion symmetry,
the many-body Berry phase defined via the Resta formula~\cite{Resta1998}
\[
\gamma = \mathrm{Im}\,\log
\langle \Psi | e^{ i \frac{2\pi}{L}\hat X } | \Psi \rangle
\]
is quantized to $0$ or $\pi$ (mod $2\pi$).
Such phases belong to a $\mathbb{Z}_2$ symmetry-protected topological (SPT) class
\cite{Schnyder2008,Pollmann2012,MorimotoFurusakiMudry2015}.
As long as the bulk gap does not close and the protecting symmetry
is preserved, all Hamiltonians within the same adiabatic class
share the same quantized Berry phase.
In one dimension, this SPT phase admits a free (quadratic) representative
without breaking symmetry
\cite{MorimotoFurusakiMudry2015}.

In the present SSH--Hubbard model,
the density is completely insensitive to the flux parameter $\theta$,
and inversion symmetry restricts the KS potential to remain
adiabatically connected to this free SSH-type representative.
Because the interacting ground state also lies in the same
$\mathbb{Z}_2$ SPT class,
symmetry protection fixes the Berry phase defined over the $\theta:0\to2\pi$ cycle:
it remains unchanged under tuning $U$ as long as the bulk gap stays open and inversion symmetry is preserved.
The equality of the Berry phase in the interacting and KS descriptions
is therefore enforced by the structure of the symmetry-protected phase itself,
rather than by a density-based reconstruction of the geometric information.

At this point, one might object that,
since the Berry phase can be computed directly
from the many-body wave function,
its agreement with the KS description may appear unsurprising,
especially given that in the present SSH--Hubbard model
the twist $\theta$ enters only through the hopping sector.
One might therefore expect that a density-matching KS reference
would generically reproduce the $\theta$-geometry,
even when the KS Hamiltonian is essentially $U$-independent.

To emphasize the limitation of density-only constraints, we broaden our scope beyond the present SSH--Hubbard setting and consider a wider class of effective lattice Hamiltonians.
A typical example is a model in which the interaction itself acquires an explicit flux dependence, $U\to U(\theta)$.
In such a case, the Hamiltonian reads
\[
\hat H(\theta)=\hat H_{\rm hop}(\theta)+U(\theta)\sum_i \hat n_{i\uparrow}\hat n_{i\downarrow},
\]
where $\hat H_{\mathrm{hop}}(\theta)$ denotes the flux-inserted hopping term.
The flux-conjugate generator then contains an interaction-induced contribution,
\[
\hat J(\theta)\equiv -\partial_\theta \hat H(\theta)
= -\partial_\theta \hat H_{\rm hop}(\theta)
-\bigl(\partial_\theta U(\theta)\bigr)\sum_i \hat n_{i\uparrow}\hat n_{i\downarrow}.
\]
The interaction-induced term can in principle compensate the local charge response from the hopping sector,
so that the density profile remains strictly $\theta$-independent along the cycle,
while the many-body holonomy remains nontrivial.
The above example explicitly shows that density constraints do not,
in general, determine the global Berry-phase geometry.

This demonstrates that the symmetry-enforced agreement observed
in the present model is not structurally protected
by density matching itself.

The present SSH--Hubbard results thus illustrate a symmetry-enforced agreement,
rather than a universal principle relating density matching to geometric equivalence.
Even more strikingly, in the strong-coupling regime
the quantum metric itself becomes strongly suppressed,
reflecting the freezing of charge fluctuations,
while the Berry phase remains symmetry-quantized.
This separation between local geometric response
and global topological holonomy highlights a structural hierarchy:
local one-point observables are insensitive to topology,
and even local geometric indicators may become inert at large $U$,
whereas the global Berry phase remains protected.
In the present symmetry-protected setting,
the KS--many-body agreement of Berry phases is therefore enforced by topological quantization itself,
rather than by density-based reconstruction.

A natural next step is to clarify how special the present symmetry is.
First, it is important to test regimes without the protecting symmetry, where the Berry phase is no longer quantized:
even if a KS construction reproduces the density exactly, it is not obvious whether the associated Berry phase can remain stable, or whether it becomes generically unconstrained.
Second, it is interesting to ask whether similar KS--many-body agreement can persist in phases protected by other interacting classifications (e.g., $\mathbb{Z}_8$ reductions), where symmetry protection constrains the adiabatic class in a different way.
Finally, these comparisons would directly address whether the SSH--Hubbard case studied here is representative of a broader mechanism (symmetry-enforced sector matching) or instead an exceptional coincidence tied to the specific symmetry and density-flat setting.

%% file: sections/A_appendix.tex
\section{Noninteracting limit and Bloch Hamiltonian}
\label{app:bloch}
In the noninteracting limit $U=0$, the model reduces to a
spin-degenerate SSH chain.
Because the system is periodic with a two-site unit cell,
Bloch's theorem allows us to express the single-particle
Hamiltonian in momentum space within the $(A,B)$ sublattice basis.

The resulting $2\times2$ Bloch Hamiltonian reads
\begin{equation}
\label{eq:Bloch_H}
\mathcal{H}(k)
=
-
\begin{pmatrix}
0 & t_1 + t_2 e^{-\mathrm{i}k} \\
t_1 + t_2 e^{\mathrm{i}k} & 0
\end{pmatrix},
\end{equation}
where
$t_1 = t(1-\delta)$ and
$t_2 = t(1+\delta)$.
Equivalently, one may write
$\mathcal{H}(k)=\vec{d}(k)\cdot\vec{\sigma}$
with
\[
d_x(k)=-(t_1+t_2\cos k),\quad
d_y(k)=-(t_2\sin k),\quad
d_z(k)=0.
\]
Here, we use the Pauli matrices in the $(A,B)$ sublattice basis,
\[
\vec{\sigma}=(\sigma_x,\ \sigma_y,\ \sigma_z),
\]
where
\[
\sigma_x=\begin{pmatrix}0&1\\1&0\end{pmatrix},\quad
\sigma_y=\begin{pmatrix}0&-\mathrm{i}\\ \mathrm{i}&0\end{pmatrix},\quad
\sigma_z=\begin{pmatrix}1&0\\0&-1\end{pmatrix}.
\]
The single-particle energies are
\begin{equation}
\varepsilon_{\pm}(k)
=
\pm \sqrt{t_1^2+t_2^2+2t_1 t_2 \cos k},
\end{equation}
which are doubly degenerate due to spin.
The band gap is therefore
\begin{equation}
\Delta_{U=0}
=
2|t_2-t_1|
=
4t\delta,
\end{equation}
so that the dimerization parameter $\delta$ directly controls
the excitation gap.

A normalized eigenvector for the lower band is
\begin{equation}
|u_-(k)\rangle
=
\frac{1}{\sqrt{2}}
\begin{pmatrix}
e^{-\mathrm{i}\varphi(k)} \\
1
\end{pmatrix},
\qquad
e^{\mathrm{i}\varphi(k)}
=
\frac{t_1+t_2 e^{\mathrm{i}k}}
{|t_1+t_2 e^{\mathrm{i}k}|}.
\end{equation}

The Berry phase of the occupied band is
\[
\gamma
=
\frac{1}{2}\int_{0}^{2\pi}\mathrm{d}k\,\partial_k\varphi(k)
\quad (\mathrm{mod}\ 2\pi).
\]
This phase equals $\pi$ if the complex function
$t_1+t_2 e^{\mathrm{i}k}$ winds once around the origin
in the complex plane, and vanishes otherwise.
For the present convention, this yields
\[
\gamma=
\begin{cases}
\pi \ (\mathrm{mod}\ 2\pi), & |t_2|>|t_1|, \\
0 \ (\mathrm{mod}\ 2\pi), & |t_2|<|t_1|.
\end{cases}
\]

\section{List of U values in this work}
\label{app:strength_U}
The many-body (MB) calculations are performed on a nonuniform $U$ grid consisting of the baseline set
{\small
  \[
  U=[0.0,0.3,0.5,0.7,1.0,2.0,3.0,4.0,6.0,8.0,10.0,15.0,20.0],
  \]
  }
augmented by additional points in the intermediate-$U$ regime:
{\small
\[
U=[0.6,\,0.9,\,1.2,\,1.5,\,1.8,\,2.1,\,2.4,\,2.5,\,2.7,\,3.5,\,4.5,\,5.0].
\]
}
The Kohn--Sham (KS) results are evaluated only on the baseline set.